\documentclass[aps,prl,twocolumn,latexsym,showpacs,subfigure,superscriptaddress,nobibnotes,nofootinbib]{revtex4-1}
\usepackage[dvips]{graphicx}
\usepackage[table]{xcolor}
\usepackage{epsfig}
\usepackage{bm}   
\usepackage{dcolumn}
\usepackage{graphicx}
\usepackage{rotate}
\usepackage{amsfonts}

\begin{document}

\title{Inference of weak nuclear collectivity from atomic masses}

\author{J. N.~Orce}
\email{jnorce@triumf.ca} \homepage{http://www.pa.uky.edu/~jnorce}
\affiliation{TRIUMF, 4004 Wesbrook Mall, Vancouver, BC V6T 2A3,
Canada}

\date{\today}

\begin{abstract}

We explore weakly-collective singly-closed shell nuclei with high-j shells where active valence neutrons  
and particle-particle correlations may be the dominant collective degree of freedom. 
The combination of large and close-lying proton and neutron pairing gaps extracted from experimental masses 
seems to charaterize the origin of the weak collectivity observed in Ni and Sn superfluids with $N\approx Z$.  
The trend of $E2$ transition strengths, i.e., $B(E2; 2^+_1\rightarrow 0^+_1)$ values, in these nuclei is predicted from proton and 
neutron pairing-gap information. The agreement with the Ni isotopes is excellent and recent experimental results 
support the trend in the Sn isotopes. This work emphasizes the importance of atomic masses in elucidating nuclear-structure properties. 
In particular, it indicates that many-body microscopic properties such as 
nuclear collectivity could be directly inferred from  more macroscopic average properties such as atomic masses. 

\end{abstract}



\maketitle

Nuclear collectivity is controlled by the interplay of particle-hole ($ph$) and 
particle-particle ($pp$) excitations. Particle-hole correlations  produce deformation
through the proton-neutron ($pn$) interaction and give rise to nuclear rotations~\cite{sc}. 
In this work, we search for weakly-collective low-lying structures in $N\approx Z$ nuclei 
where $pp$ correlations may, a priori, be the dominant 
degree of freedom. 
Large separation energies between
single-particle orbits due to the spin-orbit interaction~\cite{sm},
together with the attractive short-range pairing interaction 
acting on $J=0$ Cooper pairs~\cite{bcs} should lead to
specially stable and spherical nuclei. 
Singly-closed shell Ni ($Z=28$) and Sn ($Z=50$) isotopes with $N \gtrsim 28$ and $N \gtrsim 50$ are characterized by weakly-collective 
reduced transition probabilities, i.e., $B(E2; 2^+_1\rightarrow 0^+_1)$ values. 
Moreover, small quadrupole moments of $Q_S(2^+_1)\approx 0.05$ eb have been determined 
for $^{60-62}$Ni and $^{112}$Sn from reorientation-effect measurements~\cite{qmoments}. 
Quadrupole collectivity 
cannot solely arise from valence neutrons and proton-core excitations are needed 
to account for such weakly-collective systems. Proton-core excitations are 
supported by the positive {\it g factors} of 2$^+_1$ states in the even-mass
$^{58-64}$Ni isotopes~\cite{kenn} and the enhancement of $B(E2;
2^+_1\rightarrow 0^+_1)$ values  observed in the neutron-deficient Sn isotopes
 as the $N=50$ shell closure is approached~\cite{108sn,110sn,110sn2,sn_msu,112sn,kumar}.

The latter unveils one of the major conflicts encountered by the nuclear shell model ($SM$). 
Plainly, large-scale $SM$ calculations 
predict an inverse parabolic  trend of $B(E2; 2^+_1\rightarrow 0^+_1)$ 
values peaking at midshell and  cannot reproduce the enhancement of $E2$
strengths determined in the  $^{106-112}$Sn isotopes using $^{88}$Sr, $^{90}$Zr or $^{100}$Sn cores~\cite{108sn,110sn}. 
The former cores provide better results and 
support proton-core excitations. 
A similarly baffling scenario has recently 
been revealed by Jungclaus and collaborators at midshell of the tin isotopic chain~\cite{andrea}. 
High-statistics Coulomb-excitation measurements in inverse kinematics and fits to 
lineshapes have provided very accurate lifetimes for the  2$^+_1$ states in the $^{112,114,116}$Sn isotopes. 
Longer lifetimes  from the accepted values in the nuclear data evaluation~\cite{nndc} yield 
$B(E2; 2^+_1\rightarrow 0^+_1)$ values which clearly deviate from the inverse parabolic trend at midshell and, 
instead, propose a conspicuous minimum at $^{116}$Sn; in agreement with $N=66$ being a subshell closure.

%

\begin{figure}[!h]
\begin{center}
\includegraphics[width=6.5cm,height=5cm,angle=-0]{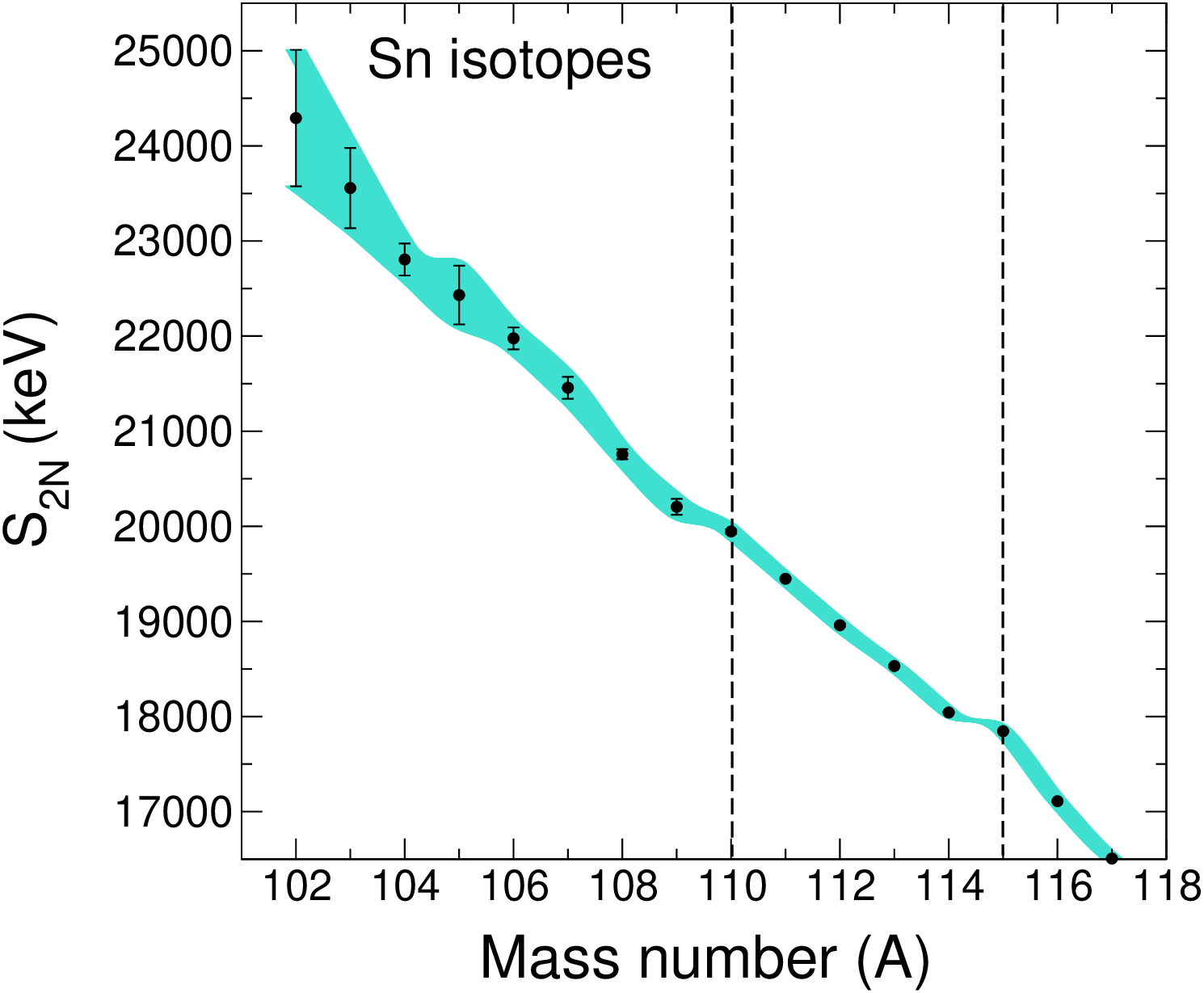}
\caption{(Color online) Two-neutron separation energies for the $^{102-117}$Sn isotopes. 
Small deviations from the smooth trend  arise at $^{110}$Sn 
and $^{115}$Sn.  A cubic-spline interpolation has been used for visual purposes.}
\label{fig:s2n}
\end{center}
\end{figure}

\begin{figure*}[t]
\begin{center}
\includegraphics[width=6.3cm,height=6cm,angle=-0]{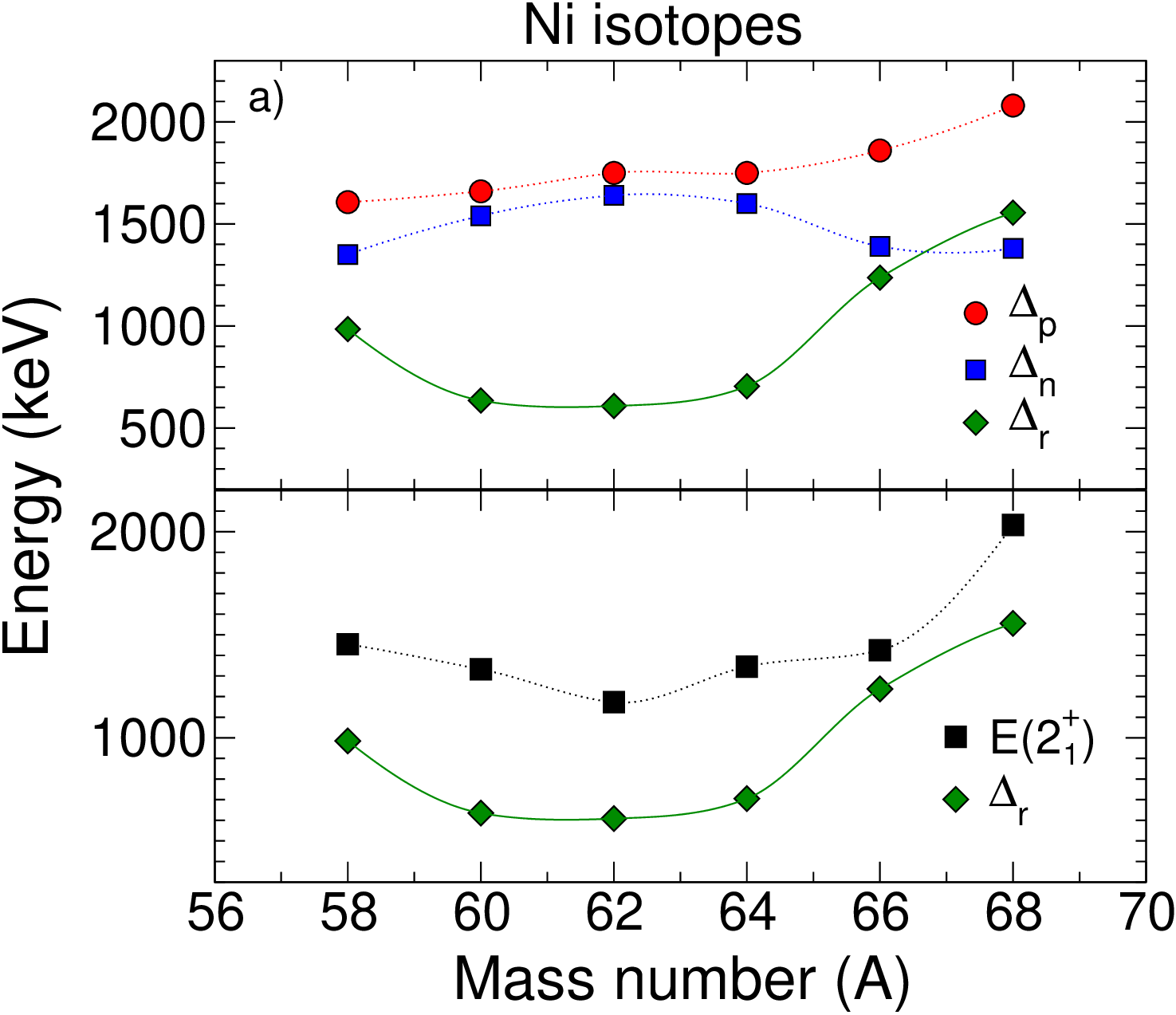}
\hspace{0.4cm}
\includegraphics[width=6.cm,height=6cm,angle=-0]{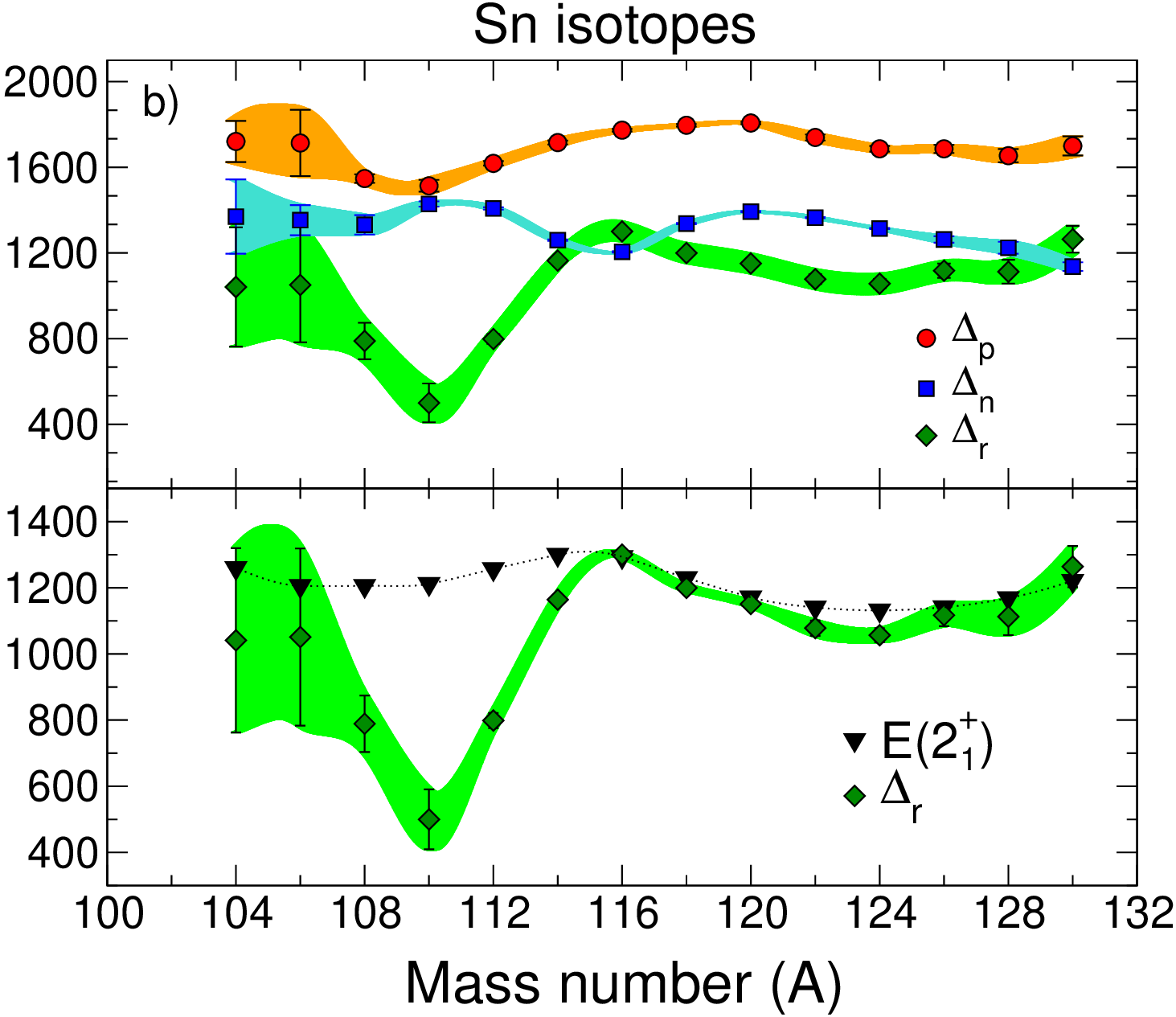}
\caption{(Color online) The top panel shows proton,  $\Delta_p$, neutron, $\Delta_n$ and relative $\Delta_r$,  pairing gaps 
extracted from the 2003 Atomic Mass Evaluation (AME03)~\cite{audi} 
for the even-mass Ni (left) and Sn (right) isotopes. The bottom panel shows a comparison 
of 2$^+_1$ excitation energies and $\Delta_r$ values. A cubic-spline interpolation has been used to remark uncertainty effects of $\Delta_r$ values in the Sn isotopes. 
Mass uncertainties regarding pairing gaps in the Ni isotopes are considered negligible, although this is not the case for $^{68}$Ni.}
\label{fig:3}
\end{center}
\end{figure*}

Furthermore, average nuclear properties such as charge radii and quadrupole deformations are directly related, through a density function, 
to  ground-state masses~\cite{deformation,thomasfermi,hfbcs,hfbcs2}. 
Similarly, nuclear masses 
provide a sensitive indicator, through  binding energies,  
for  structural changes within an isotopic chain. 
A beautiful example is given by the high-precision mass measurements of neutron-rich Mo, Zr and Sr isotopes~\cite{hager}, 
where a major change in nuclear structure is deduced from the smooth trend in two-neutron separation energies, S$_{2N}$. 
An onset of deformation at $N=60$ for $^{100}$Zr and $^{98}$Sr, as compared with much weaker deformations 
in more neutron-deficient isotopes,  is characterized by a sudden drop in the binding energies. 
Shape coexistence has been observed and 
strong rotational bands  built on the ground state and low-lying 0$^+_2$ excitations in  
$^{100}$Zr and $^{98}$Sr~\cite{nndc}. 
Further high-precision mass measurements of neutron-rich Sn isotopes 
advocates for a restoration of the  $N=82$ shell closure~\cite{masssn,132sn}.  
Figure~\ref{fig:s2n} shows S$_{2N}$ values from $^{102}$Sn to $^{117}$Sn. 
Small deviations from the 
smooth trend at $^{110}$Sn and $^{115}$Sn may suggest structural changes. 
The latter points at the $N=66$ subshell gap as supported by the minimum in the collective trend at $^{116}$Sn~\cite{andrea}. The origin of the small deviation 
at $^{110}$Sn is  more obscure. 
This work attempts at elucidating whether weakly-collective $B(E2; 2^+_1\rightarrow 0^+_1)$ values
in the Ni and Sn isotopes and atomic masses are  related in a comprehensive manner. 

Within the BCS pairing model~\cite{bcs} the 2$^+_1$ excitation is created by breaking one Cooper pair,  
and is interpreted as a two quasiparticle which lies at least twice the pairing energy, $2\Delta$. 
For  $N\approx Z$ nuclei, proton and neutron Fermi surfaces lie close to each other, 
henceforth we assume that the interplay of both proton and neutron pairing gaps may contribute 
to the overall oscillation of the Fermi surface and the collective origin of the 
2$^+_1$ state~\cite{b,ipv,broglia}. Intuitively, we 
introduce the {\it relative  pairing gap},
$\Delta_r$, defined by,
\begin{equation}
\Delta_r^2 ~ \equiv ~\mid (\Delta_p -
\Delta_n) (\Delta_p + \Delta_n) \mid ~= ~\mid \Delta_p^2 - \Delta_n^2 \mid \label{eq:deltar},
\end{equation}
\noindent  where the first term ($\Delta_p -
\Delta_n$) is the {\emph resonant factor}, which accounts for the
proximity of proton and neutron pairing-gaps. That is, the
smaller the energy difference between both pairing gaps, the larger the overlap of 
proton and neutron pairing fields. The second term  ($\Delta_p + \Delta_n$) is the {\emph
energy factor}, and accounts for the energy that can be provided to
the nuclear system before breaking Cooper pairs, i.e., a quantity that 
enhances the possibility of having spherical nuclei, where vibrations may occur.

\begin{figure*}
\begin{center}
\includegraphics[width=6.3cm,height=5.2cm,angle=-0]{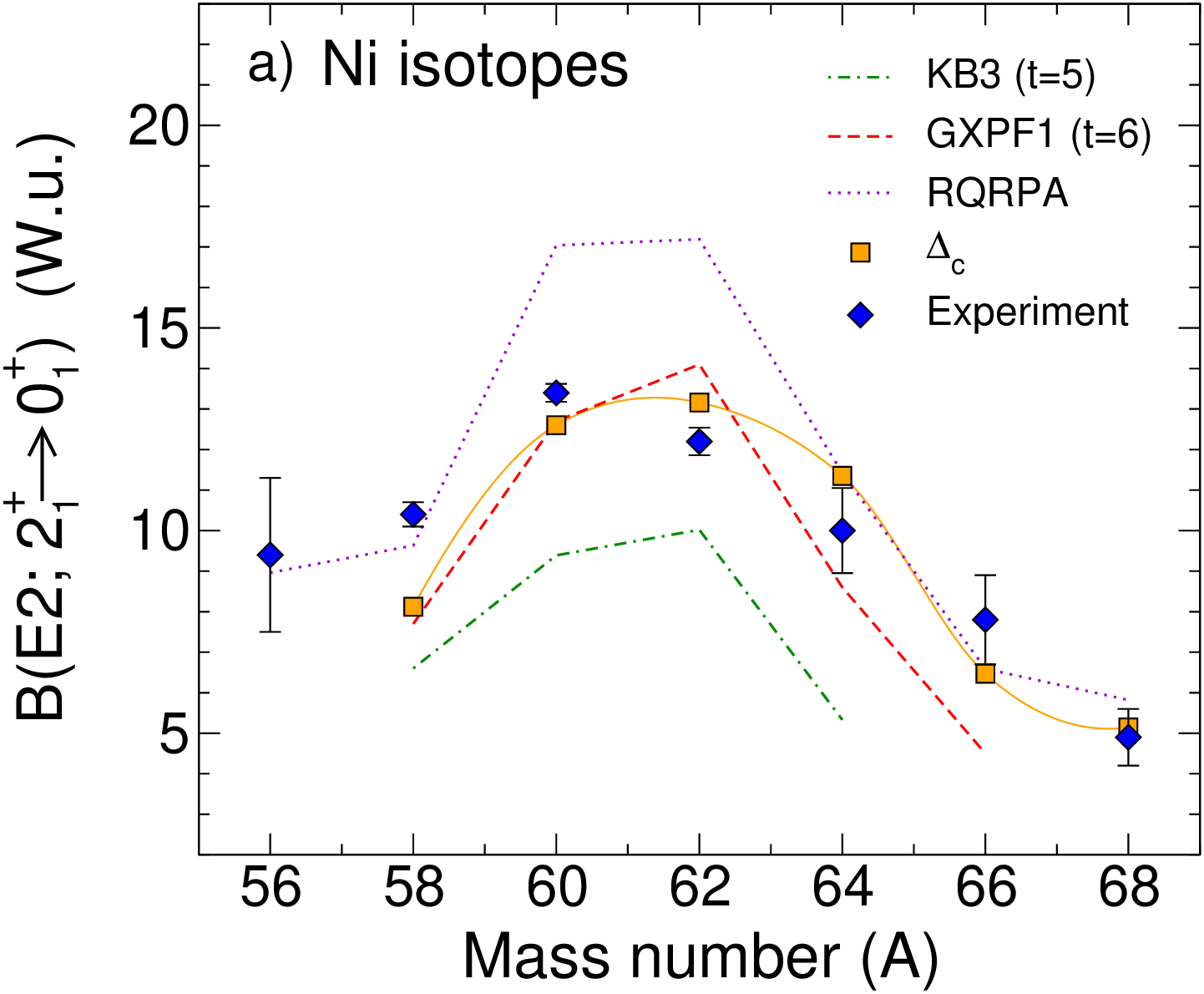}
\hspace{0.3cm}
\includegraphics[width=6cm,height=5.2cm,angle=-0]{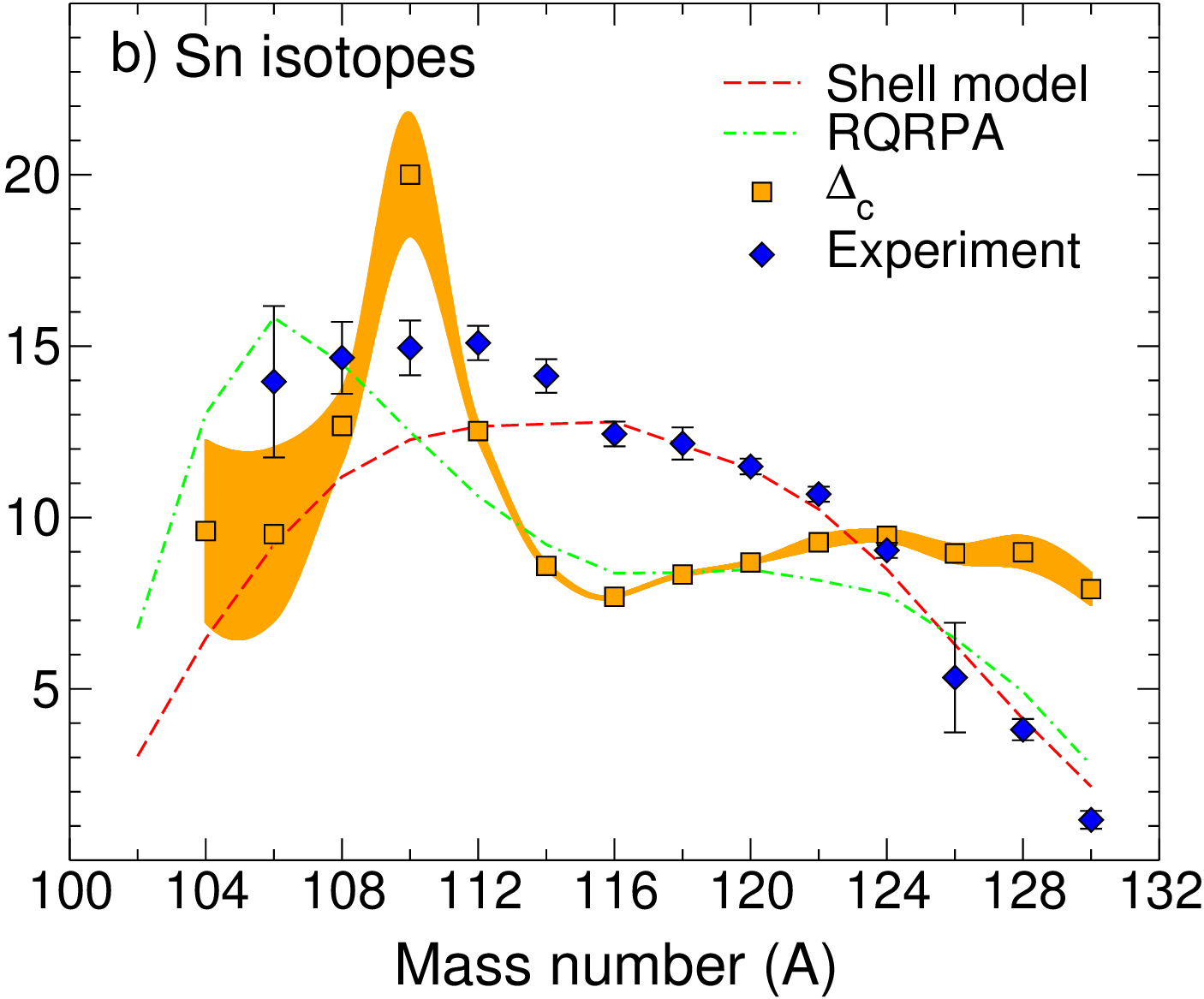}
\caption{(Color online) Experimental $B(E2; 2^+_1\rightarrow 0^+_1)$ 
weighted averages~\cite{kenn,58ni_orce,112sn,108sn,110sn,110sn2,sn_msu,112sn,kumar,raman} and $\Delta_c$
values in even-mass  a) Ni and b) Sn  isotopes.
$\Delta_c$ values are given in  units of MeV$^{-1}$.
Available $MF$ (RQRPA)~\cite{ansariring} and
large-scale $SM$~\cite{honma0,108sn} calculations are shown for
comparison. The quantity $t$ refers to the number of 
nucleons that are excited from the $f_{7/2}$ orbit to the remaining $fp$ shell using either the 
KB3 or GXPF1 effective interactions. 
A cubic-spline interpolation has been used to remark uncertainty effects of $\Delta_r$ values in the Sn isotopes. 
Mass uncertainties regarding pairing gaps in the Ni isotopes are considered negligible.} \label{fig:be2s}
\end{center}
\end{figure*}

%
The 
magnitude of the neutron, $\Delta_n$, and proton, $\Delta_p$,
pairing gaps can be determined from experimental odd-even mass differences~\cite{audi} derived from the 
Taylor expansion of the nuclear mass in nucleon-number 
differences~\cite{deformation}. 
These prescriptions assume that pairing is the only non-smooth
contribution to nuclear masses. We extract $\Delta_n$ and $\Delta_p$ 
from the symmetric five-point difference~\cite{deformation},  
which accounts better for  blocking effects in odd-mass nuclei between shell gaps~\cite{bender},
\begin{eqnarray}
\Delta_n^{(5)} &=& -\frac{1}{8} [M(Z,N+2)-4M(Z,N+1)+6M(Z,N) \nonumber  \\ 
&-& 4M(Z,N-1)+M(Z,N-2) ]\\
\Delta_p^{(5)} &=& -\frac{1}{8} [M(Z+2,N)-4M(Z+1,N)+6M(Z,N) \nonumber  \\
&-& 4M(Z-1,N)+M(Z-2,N) ].
\end{eqnarray}  
\noindent Here, we make the strong assumption of a valid $\Delta_p$ in the region of study, 
although the kink of binding energies at shell closures would, a priori, not allow a Taylor expansion. 
This assumption is supported by the $^{56}$Ni and $^{100}$Sn soft cores and Ref.~\cite{bender}. 
Doubly magic $^{56}$Ni and $^{100}$Sn are not included in the pairing-gap systematics 
since in these cases both magic-number and Wigner cusps span
singularities in the mass surface~\cite{madlandnix}. 
The top panel of Fig.~\ref{fig:3} shows $\Delta_n$ and
$\Delta_p$ in the even-mass Ni (left panel) and Sn (right panel) isotopes. As expected, $\Delta_n$
lies lower than the corresponding $\Delta_p$.





For comparison, $\Delta_r$ values and 2$^+_1$ excitation energies in the 
$^{58-68}$Ni  and $^{104-130}$Sn isotopes are plotted in the bottom panels
of Fig.~\ref{fig:3} (left and right panels, respectively). Excitation energies and $\Delta_r$ values
follow a similar  trend at $^{60-64}$Ni and differ for $^{58}$Ni and $^{66-68}$Ni.
For instance, whereas the energy difference of the 2$^+_1$ states in
$^{58}$Ni and $^{66}$Ni is only  $\sim$ 30 keV, there is a sharper
energy difference of 150 keV between their $\Delta_r$ values. 
$\Delta_r= 0.985$ MeV for $^{58}$Ni and it decreases
to minimum values of $\Delta_r=0.635$ and 0.608 MeV for $^{60}$Ni and for
$^{62}$Ni, respectively. 
Protons and neutron pairing gaps  begin to diverge at $^{64}$Ni, $\Delta_r=0.705$
MeV, with $\Delta_n$ becoming much smaller than $\Delta_p$.
$\Delta_r=1.237$ MeV in $^{66}$Ni, and has a maximum value of
$\Delta_r=1.555$ MeV at $^{68}$Ni, where $\Delta_p$ has a maximum
energy of $\sim$2.1 MeV and $\Delta_p - \Delta_n$ has the largest
energy difference.  


Moreover, $\Delta_r$ values and 2$^+_1$ energies in the Sn isotopes
follow a similar  parabolic trend from $^{116}$Sn to $^{130}$Sn.
However, unlike excitation energies, the trend of $\Delta_r$ values
between $^{104}$Sn to $^{116}$Sn clearly shows a sharp minimum at
$^{110}$Sn with $\Delta_r=0.498$ MeV. For lighter and heavier Sn isotopes, proton and
neutron pairing gaps begin to diverge, with $\Delta_r$ values of
0.791 and 0.799 MeV at $^{108}$Sn and $^{112}$Sn, respectively.
$\Delta_r$ increases to a maximum value of 1.300 MeV for $^{116}$Sn.
From $^{116}$Sn to $^{130}$Sn,  $\Delta_r$ values follow a smooth
parabolic trend,  with $\Delta_r$ being  larger than for
$^{108-112}$Sn.  Larger $\Delta_r$ values of 1.076 and 1.010 MeV are found for $^{104}$Sn and
$^{106}$Sn, respectively.
A strong correlation with  quadrupole collectivity can be inferred from the trends of $\Delta_r$ values  in the Ni and Sn isotopes. 
%


%
%
%

From a global fit to available $B(E2; 0^+_1 \rightarrow 2^+_1)$ values 
throughout the nuclear chart, Grodzins deduced an exceptional formula that calculates  
surprisingly well $B(E2; 0^+_1 \rightarrow 2^+_1)$ values from well-known 2$^+_1$ 
energies~\cite{grodzins}.  Raman improved 
the fit from a larger data set~\cite{raman} and 
the Grodzins-Raman's empirical formula is given by, 
\begin{eqnarray}
B(E2; 0^+_1 \rightarrow 2^+_1)  =  \left( 2.57 \pm 0.45 \right) Z^2A^{-2/3}E(2^+_1)^{-1}
\end{eqnarray}
The physical meaning of this formula remains unknown. 
\noindent Similarly, given the qualitative agreement between $\Delta_r$ values and 2$^+_1$ energies 
in the even-mass Ni and Sn isotopes, $E2$ collectivity might  be
estimated using the inverse of
$\Delta_r$. For that,  the {\it pairing-gap collective strength}, $\Delta_c$,
is defined as,
\begin{equation}
\Delta_c \equiv \frac{2\Omega}{\Delta_r},
\label{eq:deltac}
\end{equation}
\noindent where $2\Omega=(2j+1)$ is the average particle number, i.e., 
the total number of proton and neutron Cooper pairs that may contribute to the 
collective motion. $\Delta_c$ values are given in  units of $[E^{-1}]$. 
In order to examine the interplay of proton-core excitations and $pp$ correlations 
for Ni and Sn isotopes with $N\approx Z$, only the   f$_{7/2}$ and g$_{9/2}$  proton and neutron orbits, respectively, 
will be included in Eq.~\ref{eq:deltac}. 
These orbits are fully occupied and the large spacial
overlap of magnetic substates  may enhance pairing
correlations. This assumption may not be valid for the very neutron-rich Sn isotopes.

The origin of Eq.~\ref{eq:deltac} lies within  the BCS framework.  
For the special case of a pure pairing force in a  single-j  shell, 
the gap equation yields the two-quasiparticle energy~\cite{rs,lbcs},  
\begin{equation}
E_k + E_{k^{\prime}} = 2\Delta = G~\Omega,
\label{eq:pg}
\end{equation}
where $E_k, E_{k^{\prime}}$ are the quasiparticle energies at the Fermi surface 
and $G$ the pairing strength. 
Given the qualitative agreement between $\Delta_r$ and 2$^+_1$ energies, it can be assumed that  
$2\Delta \approx \Delta_r$ and  Eq.~\ref{eq:deltac} can be written as, 
\begin{equation}
\Delta_c \approx \frac{2}{G}.  
\label{eq:wow}
\end{equation}
That is, nuclear collectivity is inversely proportional to the pairing strength. 

Finally, Fig.~\ref{fig:be2s}  shows the systematics of experimental $B(E2;
2^+_1\rightarrow 0^+_1)$ values (blue diamonds) as compared with
single-particle estimates (1 W.u.) in the even-mass $^{58-68}$Ni~\cite{kenn,raman,58ni_orce} and $^{106-130}$Sn
isotopes~\cite{108sn,110sn,sn_msu}. 
Strikingly, the  trend of $E2$
strengths in these Ni isotopes is in  agreement
with $\Delta_c$ values (left panel of Fig.~\ref{fig:be2s}).  In fact, the trend of $\Delta_c$ values provides a 
better agreement than large-scale $SM$~\cite{honma0} and $MF$~\cite{ansariring} calculations. 
For $^{62}$Ni, $\Delta_c$ presents a maximum in the systematics which corresponds to the lowest 2$^+_1$ energy 
and the strongest collectivity predicted by large-scale $SM$~\cite{honma0}
and $MF$~\cite{ansariring} calculations.  

The  trend of quadrupole collectivity in the Sn isotopes is not as precisely defined as in the Ni isotopes, although the
enhancement of collectivity in the neutron-deficient Sn isotopes is well established.  
The trend of $\Delta_c$ values in the Sn isotopes is plotted in
the right panel of Fig.~\ref{fig:be2s} and shows an enhancement of
$E2$ strengths in the neutron-deficient Sn isotopes, with a sharp
maximum at $^{110}$Sn.  $MF$ calculations also indicate a sharp
maximum in the trend of $E2$ strengths~\cite{ansariring}, but
peaking at $^{106}$Sn. This maximum is unlikely since the energy spectrum of $^{106}$Sn shows typical properties of singly
closed-shell nuclei that can simply be explained with a $\delta$-function interaction. 
The $B(E2; 2^+_1\rightarrow 0^+_1)$ values
for $^{104}$Sn and $^{106}$Sn are, respectively, either unknown or
with large uncertainties. The most precise $B(E2; 2^+_1\rightarrow 0^+_1)$ in $^{106}$Sn by
Ekstr\"om and collaborators indicates, however, a decreasing trend with a smaller $B(E2; 2^+_1\rightarrow
0^+_1)=13.1(2.6)$ W.u. as compared with $^{108}$Sn~\cite{110sn2}, in agreement with the
trend proposed in this work. 
In addition, Jungclaus and co-workers have recently determined much lower and precise 
$E2$ strengths for $^{112,114,116}$Sn, with decreasing absolute 
$B(E2; 2^+_1\rightarrow 0^+_1)$ values  from $^{112}$Sn to $^{116}$Sn (not plotted in this work). 
These remarkable results point at $^{116}$Sn as the new minimum in the collective trend at midshell, 
in disagreement with  large-scale $SM$ calculations, and supporting the $N=66$ subshell 
closure and the current work. Summarizing, this work proposes that microscopic many-body properties such as 
nuclear collectivity could be inferred from atomic masses.

Further transition strengths and  mass measurements are needed in the neutron-deficient 
and mid-shell Sn region  to confirm the collective trend proposed in this work.
In particular, more accurate experimental data is needed in the key $^{110}$Sn isotope. 
Curiously, the experimental masses accepted in the 2003 atomic mass evaluation   
concerning pairing gaps in $^{110}$Sn are  either from unpublished private communications 
or based on $\beta$ end-point measurements~\cite{audi}.

The author gratefully  acknowledges fruitful discussions with U.
Hesse-Orce,  S. Baroni, M. Brodeur, M. Honma, A. Ansari, C. J. Lister, B. Singh and S. W. Yates. 
TRIUMF receives federal funding via a contribution agreement 
through the National Research Council of Canada.

\end{document}